\documentclass[fleqn,10pt]{wlscirep}
\usepackage[utf8]{inputenc}
\usepackage[T1]{fontenc}
\usepackage{lineno}
\usepackage{ulem}% to strike through text / remove at the end
\usepackage{placeins}% float barrier to fix figures and tables
\usepackage{setspace}
\usepackage{enumitem}
\usepackage{multirow}
\usepackage{longtable}   % put this in your preamble
\usepackage{booktabs}    % optional: nicer horizontal rules
\usepackage{xurl}
\usepackage{doi}
\hypersetup{breaklinks=true}

\title{Geo-Disasters: geocoding climate-related events in the international disaster database EM-DAT}

\author[1,*]{Khalil Teber}
\author[3]{Mélanie Weynants}
\author[3]{Fabian Gans}
\author[1,2]{Miguel D. Mahecha}
\affil[1]{Institute for Earth System Science and Remote Sensing, Leipzig University, Leipzig, 04103, Germany}
\affil[2]{Center for Scalable Data Analytics and Artificial Intelligence (ScaDS.AI), Dresden/Leipzig, Germany}
\affil[3]{Max Planck Institute for Biogeochemistry, Jena, Germany}

\affil[*]{corresponding author: Khalil Teber (khalil.teber@uni-leipzig.de)}

\begin{abstract}
Climate hazards can escalate into humanitarian disasters. Understanding their trajectories---considering hazard intensity, human exposure, and societal vulnerability---is essential for effective anticipatory action. The International Disaster Database (EM-DAT) is the only freely available global resource of humanitarian disaster records. However, it lacks exact geospatial information, limiting its use for climate hazard impact research. Here, we provide geocoding of 9,217 climate-related disasters reported by EM-DAT from 1990 to 2023, along with an open, reproducible framework for updating. Our method remains accurate even when only region names are available and includes quality flags to assess reliability. The augmented EM-DAT enables integration with other geocoded data, supporting more accurate assessment of climate disaster impacts and adaptation deficits.
\end{abstract}
\begin{document}

%\linenumbers

\flushbottom
\maketitle

\thispagestyle{empty}

%\noindent Please note: Abbreviations should be introduced at the first mention in the main text – no abbreviations lists or tables should be included. Structure of the main text is provided below.

\section*{Background \& Summary}

%(700 words maximum) An overview of the study design, the assay(s) performed, and the created data, including any background information needed to put this study in the context of previous work and the literature. The section should also briefly outline the broader goals that motivated the creation of this dataset and the potential reuse value. We also encourage authors to include a figure that provides a schematic overview of the study and assay(s) design. The Background \& Summary should not include subheadings. This section and the other main body sections of the manuscript should include citations to the literature as needed.
Extreme climate events can negatively impact societies in various ways \cite{carletonSocialEconomicImpacts2016a} and can escalate into humanitarian disasters with heavy loss and damage tolls. Understanding their impact dynamics across multiple events is crucial for future disaster risk reduction efforts. Quantifying hazard intensities as well as exposure and vulnerability to them within a consistent framework is important, as these factors co-determine expected impacts \cite{ fieldManagingRisksExtreme2012a}. However, the primary challenges in this context are data quality and availability \cite{mahechaDataChallengesLimit2020}. Specifically, there is a lack of geographically explicit datasets that allow integration of climate hazards, societal exposure and vulnerability, and observed humanitarian impacts. Among these gaps, the most critical is the absence of precise disaster-impact data.

\begin{figure}[h]%
\centering
\centerline{\includegraphics[scale=0.6]{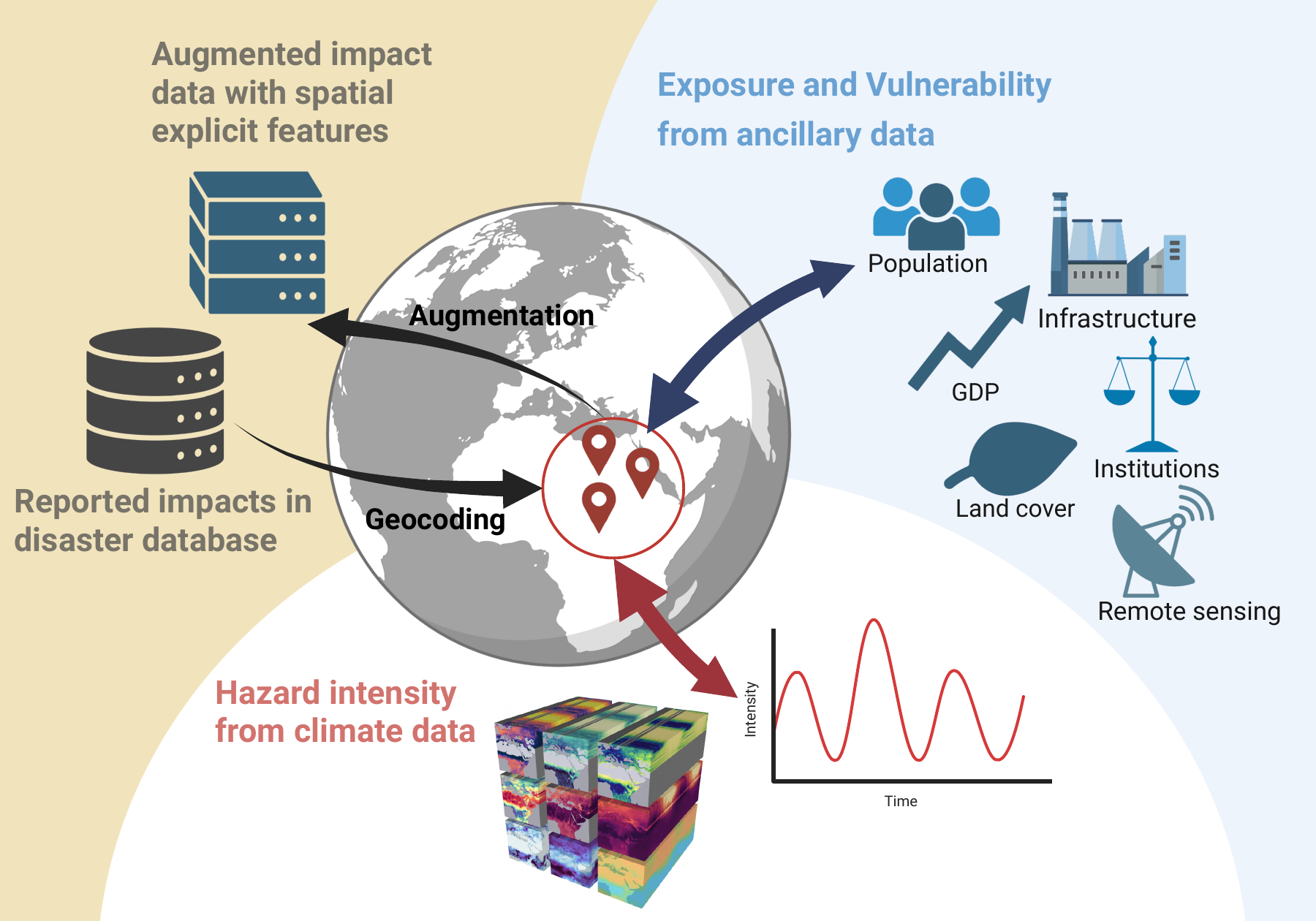}}
\caption{\textbf{Conceptual data-fusion workflow for disaster-impact analysis using geographical explicit data}. Geocoded disaster impact records provide the spatial key that unlocks joint analyses with complementary data streams. Overlaying gridded climate products yields quantitative measures of hazard intensity, while a wide range of ancillary layers characterize exposure and vulnerability. Raster-based examples include population and GDP density surfaces, land-cover maps, digital elevation and OpenStreetMap-derived infrastructure; table-based examples include sub-national indicators from the Global Data Lab, World Bank development metrics and V-Dem institutional indices. Together these linked datasets enable integrated assessments of disaster impacts and risk.
Illustration created in BioRender by Moenks, M. (2025) (https://BioRender.com/aqv9alw).}
\label{concept_figure}
\end{figure}

%HERE WE NEED ONE PARAGRAPH ON THE SOCIETAL EXPOSURE AND VULNERABILITY DATA WE WILL USE
Geographically explicit impact data unlock the potential to link disaster observations with external datasets, enabling empirical investigation of all facets of disaster risk. Fig.~\ref{concept_figure} illustrates a core concept: once disaster events are geocoded, they become the spatial reference that allows linking impact records to gridded climate fields, and a wide range of ancillary socio-economic and other datasets. Climate reanalysis data such as ERA5 \cite{hersbachERA5GlobalReanalysis2020} can be integrated to quantify hazard intensity. Potential population and economic exposure can be derived from gridded population \cite{schiavinaGHSPOPR2023AGHS2023} and Gross Domestic Product (GDP) \cite{kummuDownscaledGriddedGlobal2025}. A wealth of newly available socio-economic and political datasets---e.g., critical infrastructure \cite{nirandjanSpatiallyexplicitHarmonizedGlobal2022}, the Global Data Lab’s sub-national indicators \cite{GlobalDataLab2025}, and the V-Dem democracy indices \cite{michaelcoppedgeetal.VDemDatasetV142024}---offer insight into vulnerability. The biophysical context of impacted territories can be characterized with land cover products \cite{esaLandCoverCCI2017}, digital elevation models, and flood plain maps \cite{nardiGFPLAIN250mGlobalHighresolution2019}. Yet, all these extensions hinge on accurately identifying the regions affected by each disaster.

The Emergency Events Database (EM-DAT) \cite{emdat2025} is the most widely used global data set on humanitarian disaster impacts \cite{Delforge2025}. It is freely available and documents human losses and economic damage resulting from both human-made and natural disasters. EM-DAT has become the key resource for analyzing global records of event occurrences and their corresponding impacts \cite{donattiGlobalHotspotsClimaterelated2024, Delforge2025}. It is well known that the reporting is not bias free despite strict criteria to determine which disasters are recorded. Reasons can be found in the information management of local authorities and other issues \cite{moriyamaComparisonGlobalDatabases2018}. Although reporting quality has improved over the past two decades, many impact numbers remain estimates and contain missing values \cite{wirtzNeedDataNatural2014, jonesHumanEconomicImpacts2022, Delforge2025}. However, it remains a very valuable global data resource and thus needs to be improved. One avenue is to overcome the lack of geospatially explicit data sources which constrains its usability for spatio-temporal analysis.

Despite its well-known gaps, EM-DAT has proven to be a valuable resource for the study of the dynamics and impacts of different climate related hazards. For example, the better-geocoded portion of EM-DAT facilitated the spatially explicit analysis of flood impacts, revealing correlations between income inequality and flood fatalities \cite{linderssonWiderGapRich2023}; it has also served as a complement to the flood impact database FLODIS \cite{mesterHumanDisplacementsFatalities2023}, which was in turn used to examine global vulnerability to floods in the last two decades \cite{sauerLimitedProgressGlobal2024}. However, these examples underscore a broader point: when precise locations are available, analysts can quantify exposure and vulnerability directly; when they are not, substantial parts of EM-DAT remain out of reach. Recognizing this challenge, one initiative—the Geocoded Disasters (GDIS) extension \cite{rosvoldGDISGlobalDataset2021a}—has attempted to geocode the full EM-DAT catalogue. We turn next to that effort and evaluate how well it addresses the lack of precise geolocations.

\begin{figure}[h]%
\centering
\centerline{\includegraphics[scale=0.2]{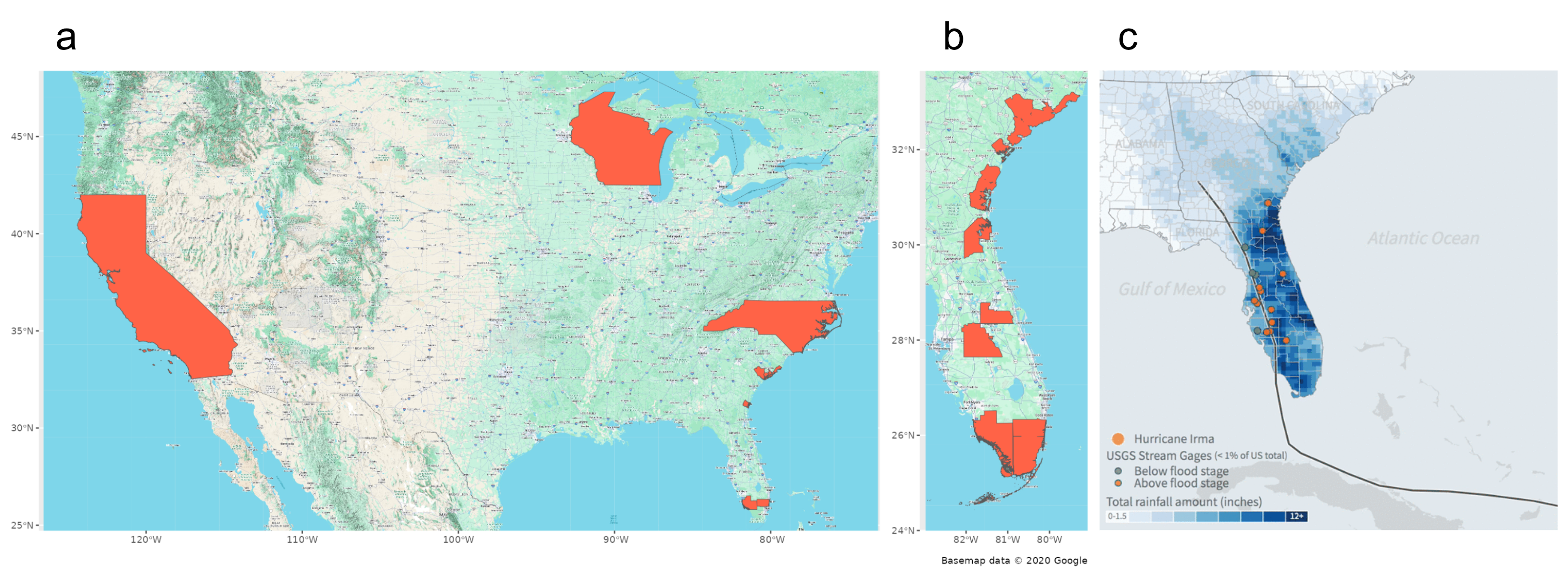}}
\caption{\textbf{An example of a geocoded disaster in Geo-Disasters vs GDIS database}. \textbf{ (a)} GDIS geocoding. \textbf{(b)} Geo-Disasters geocoding. \textbf{\textbf{(c)}} Hurricane trajectory in mainland USA according to USGS. Geocoding events is especially challenging in the United States of America, where the $\sim$ 3000 counties have $\sim$ 1900 names. Our approach that allows making use of the GAUL ID when available, or including all relevant information in the GeoNames query provides more accurate results than GDIS, as illustrated by this example.}
\label{hurricane_irma}
\end{figure}

The Geocoded Disasters (GDIS) extension provides geographically explicit geocoding of EM-DAT disaster events between 1960 and 2018, which enabled the aforementioned studies \cite{linderssonWiderGapRich2023, mesterHumanDisplacementsFatalities2023, sauerLimitedProgressGlobal2024}. However, we see several pitfalls that inform our new framework. First, because GDIS relies on the reference administrative database Global Administrative Areas (GADM) \cite{GADMDATA} rather than EM-DAT’s native Global Administrative Unit Layers (GAUL) \cite{gaul2015}, name-based matching can be vulnerable to homonyms. For example, Hurricane Irma (2017) is mapped by GDIS even to California and Wisconsin--although EM-DAT reports landfall only in Florida (Fig.~\ref{hurricane_irma}a). Second, the underlying geocoding scripts are not publicly available, precluding replication or improvement. Third, other errors in the database, such as incorrect ISO-3 identifiers for certain countries, hinder data matching with EM-DAT and raise quality concerns. Fourth, the absence of quality flags leaves users unsure whether locations were assigned automatically or manually. Finally, using administrative level 3--unsupported in EM-DAT--introduces a scale mismatch whose effect on event footprints remains unclear. Together, these shortcomings highlight the need for a new, open, reproducible, and EM-DAT–compatible geocoding framework that builds on the lessons from GDIS and can be routinely updated and validated.

In this paper, we introduce Geo-Disasters, an  openly accessible database and its accompanying geocoding framework that locates climate-related EM-DAT disasters.
We use the GAUL administrative reference database to leverage all available EM-DAT information and minimize uncertainty. We limit the geocoding to climate disasters recorded between 1990 and 2023 that include at least one impact metric. Each event is assigned a quality flag that grades the reliability of its geocoding. By expanding the pool of precisely geocoded events, Geo-Disasters and its transparent framework enable finer-scale analyses, essential for adaptation planning and disaster-risk reduction.

\section*{Methods}

Geo-Disasters provides geocoding for 9,217 unique climate-disaster events reported by EM-DAT for the period 1990-2023, together with quality flags that grade location certainty. EM-DAT lists names of regions where disaster impacts were reported for a given event. Since 2000, EM-DAT includes GAUL identification codes for most reported regions. Therefore, initiatives that rely on GADM database---an alternative to GAUL---restrict geocoding to name-based matching and fail to use valuable information embedded in EM-DAT. Such approaches are error-prone and risk introducing bias to downstream analysis. These considerations led us to adopt GAUL as the reference framework for Geo-Disasters.

\begin{figure}[h]%
\centering
\centerline{\includegraphics[width=\linewidth]{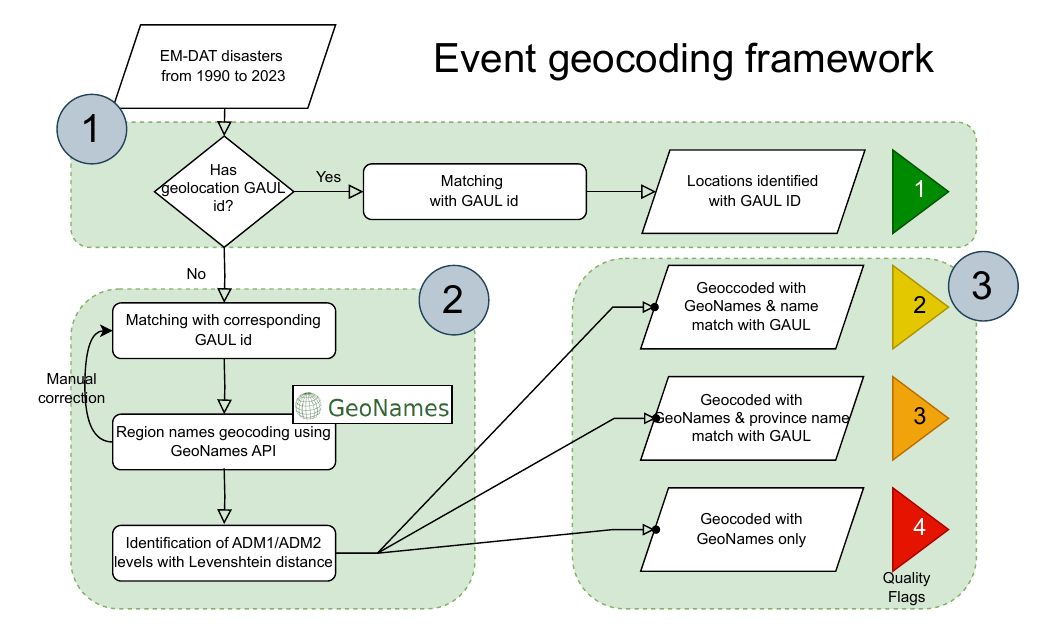}}
\caption{\textbf{Event Geocoding framework}. The majority of the events are geocoded using the GAUL ID provided by EM-DAT and have the highest quality flag. For the remaining locations, different approaches are combined to identify the different locations, including manual correction of location names. Different quality checks are combined, using the geocoding client GeoNames and matching by location names to assign a quality flag to the identified location.}
\label{geocoding_framework}
\end{figure}

Our geocoding workflow (Fig.~\ref{geocoding_framework}) proceeds in three stages: (1) GAUL-ID matching, (2) GeoNames fallback, and (3) quality flagging. In the first step, we use the GAUL identification codes attached to most post-2000 events to assign each disaster to its correct sub-national region(s).  (Fig.~\ref{geocoding_framework}, step 1). Because the GAUL identification encodes the relevant administrative level, no further action is required. All locations derived directly from a GAUL code are therefore tagged with the quality flag (1-highest). From the identified 45,121 locations, 39,225 (86\%, corresponding to 7,375 disasters) have the highest quality flag.

In the second step (Fig.~\ref{geocoding_framework}, step 2), we geocode events lacking a GAUL identifier by querying the freely available GeoNames database \cite{geonames2025}. Because these records contain only textual place names, we first standardize and clean the entries to maximize accuracy. Many names exhibit spelling variants, typographical errors, or ambiguous wording, and others include superfluous descriptors-such as "Near," "Between", "Province," or "District"-that do not directly correspond to official GeoNames labels. We therefore strip generic qualifiers, harmonize diacritics, and apply case-insensitive matching before submitting the names to GeoNames. We also correct outdated or incorrect ISO-3 codes: for example, AZO (Azores) becomes PRT (Portugal), and SRB (Serbia) becomes MNE (Montenegro) for events that occurred before Montenegro's independence. These preprocessing steps align all locations with standard administrative divisions, reducing mismatches during geocoding.

Next, the preprocessed location names are submitted to the GeoNames API. Each query restricts the search to the event's country, as given by its ISO-3 code. A successful match returns: (i) a longitude/latitude pair for the region's centroid, (ii) the region's standard GeoNames name, (iii) the name of its parent administrative unit (province/ADM1).

However, some location names remain unresolved because of spelling variations, incomplete information, or local naming variations. Manual review added 829 matches. For example, in event 1990-0001-LKA, "Nuwera Eliya" was mistakenly parsed as two separate locations ("Nuwara, Eliya"); in event 1992-0008-LBN, "South Bekaa" should be Beqaa; event 1993-9511-MRT listed "N.W. Assaba, Tagnat" which should read "Assaba, Tagant"; and event 1998-0070-BRA used "Macajai", a misspelling of "Mucajaí" in Roraima state. These manual corrections ensure that locations are recognized by GeoNames, thereby increasing geocoding accuracy and reducing inconsistencies.

For each case, the identified longitude/latitude pair is mapped to potential administrative regions at GAUL level 1 (ADM1) and level 2 (ADM2). The corresponding administrative level for each location is determined using the following approach:

\begin{enumerate}
    \item Exact name match: If an identified location name exactly matches a GAUL ADM1 or ADM2 region, we assign the corresponding level.
    \item Fuzzy match (Levenshtein distance): if there is no exact match, we calculate the Levenshtein similarity score and assign the administrative level (ADM1 or ADM2) of the closest match.
    \item GeoNames province fallback: if an identified location name corresponds to a province name in GeoNames, we classify it as ADM1.
    \item Containment check: if an identified ADM2 region is entirely contained in an ADM1 region already matched for the same event, we retain only the ADM2 region.
    \item Default to ADM2 when GAUL name missing: if name verification is impossible because GAUL lists a unit as "Administrative unit not available" (93 ADM1 regions or 2.74\%, and 1150 ADM2 regions or 3\% ), we assign the location to ADM2 by default.
\end{enumerate}

In the third step (Fig.~\ref{geocoding_framework}, step 3) we assign a four-tier quality flag to every geocoded region. These flags capture how closely each assigned location aligns with the spatial information reported in EM-DAT. Level 1 (highest) relies on the GAUL identifiers supplied by EM-DAT, while levels 2-4 describe progressively less certain matches:

\begin{enumerate}[label=Level \arabic*:, leftmargin=*]
    \item Highest Quality: location derived directly from the EM-DAT GAUL ID; virtually no risk of discrepancy.
    \item High Quality: GeoNames match whose name exactly corresponds to a GAUL region; Minimal risk of discrepancy.
    \item Medium Quality: GeoNames match based on fuzzy string similarity or a province name mapped to a GAUL ADM1 unit; moderate risk owing to potential name variations.
    \item Lowest Quality: fallback assignments that does not meet the above criteria; risk of discrepancy unknown.
\end{enumerate}

The final dataset comprises 45,121 geocoded locations corresponding to 9,217 unique climate-related disaster events in 207 countries and territories. Reporting is skewed towards first-level administrative units: 27,265 ADM1 polygons versus 17,856 ADM2 polygons. Overall, 4,638 events are described solely at ADM1 scale, 3,537 solely at ADM2, and 1,042 combine both levels. Footprint size varies strongly: 3,071 disasters strike a single administrative unit, whereas 1,026 affect 11 units or more. The widest footprint is a May 2003 U.S. tornado-outbreak sequence (event 2003-0210-USA), which touched 172 counties across 18 states. Beyond the spatial footprint of each event, the dataset also captures the full spectrum of climate-related hazards recorded in EM-DAT.

\begin{figure}[h]%
\centering
\centerline{\includegraphics[scale=0.5]{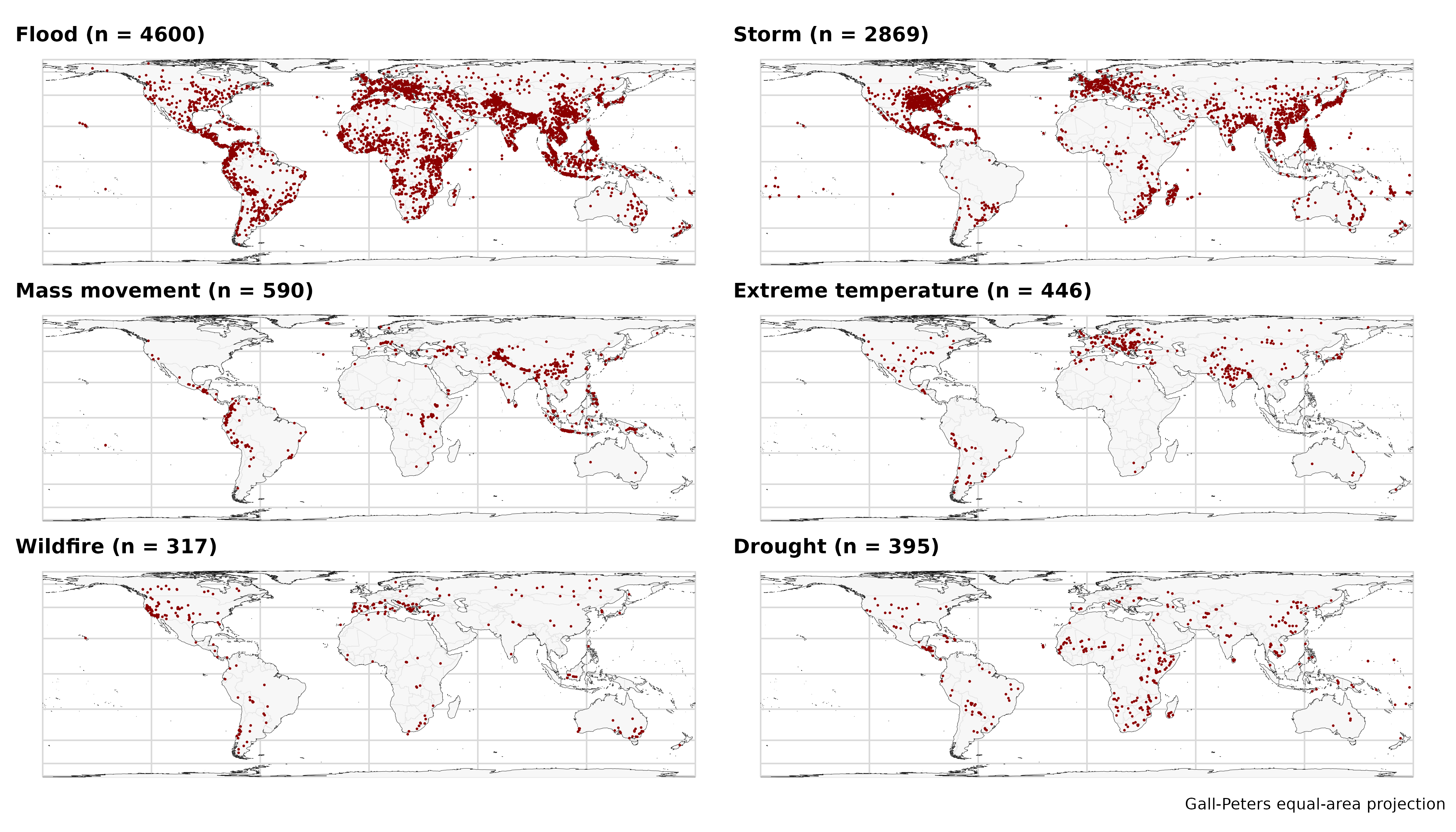}}
\caption{\textbf{Distribution of geocoded EM-DAT events by disaster type from 1990 to 2023}. The geographic distribution varies from a disaster type to the other. Floods are the most reported type of disasters and occur on all continents. Storms and mass movement are the second and third most reported disasters, and their distribution shows geographic clustering related to storm belts or typical landslide hilly / mountainous regions. On the other hand, temperature related extremes show strong latitudinal bias, especially in  the cases of extreme temperature (mostly in mid to high latitudes) or droughts (mostly reported in low latitudes).}
\label{spatial_distribution}
\end{figure}

According to the EM-DAT classification, the disasters are divided into six main hazard types: 4,600 floods, 2,869 storms, 590 mass movement (572 wet and 18 dry), 446 extreme temperature episodes (heat waves, cold waves, extreme winter conditions), 395 droughts, and 317 wildfires. Fig.~\ref{spatial_distribution} maps the spatial distribution of geocoded climate disasters by hazard type, revealing several reporting hot-spots and blind-spots. We retain only disasters useful for quantitative impact studies which have at least one non-missing impact variable (i.e., number affected, number of fatalities, or total economic damage); events lacking all three impact variables are excluded. Finally, besides providing each identified location individually, we provide a dissolved footprint per event, i.e. a single geometry that unions all component polygons.

To provide users with a single, easy-to-interpret event-level quality flag, we take the worst (highest-numbered) location flag among all regions that constitute the event. This conservative rule ensures that any uncertainty present in even one location propagates to the entire event, avoiding overconfidence in spatial accuracy. For example, a 1991 cold wave in France affected five locations--Champagne, Cognac, Provence, Touraine, and Bordeaux--where the identified regions carry two location‐level flags 2 (high quality: exact name match), one flag 3 (medium quality: fuzzy string match) and two flags 4 (low quality: no name match). Under our scheme the aggregated event footprint is labeled quality 4, signaling to users that, despite mostly precise geocoding, at least one part of the footprint is less certain. Researchers requiring the strictest accuracy can therefore filter events by quality flag $\leq$ 2, whereas those tolerant of moderate uncertainty may retain all events while still being aware of the underlying limitations. Adopting a single, worst-case event-level flag streamlines reliability reporting for most users, while the original location-level flags remain available for users who require finer control.

We carried out several additional manual verification steps to enhance the accuracy and reliability of the data. Locations with vague descriptors such as "entire country", "northern" or "western", were excluded unless they matched an official administrative unit. Events with lower quality flags (2-4) underwent a manual review, and those with poorly resolved locations were removed. Certain countries required special handling because GAUL region names and EM-DAT reported locations differ. In Burkina Faso and Haiti, for instance, GAUL records French toponyms whereas EM-DAT often lists English translations, producing mismatches. Inconsistent spellings--e.g. N’Djamena/N’djamena/N’djam for Chad's capital--were also harmonized. Language mismatches affected other records as well, notably the USA and Haiti, where some events are reported in French. The United States posed a unique challenge: 3,144 counties (ADM2) include 428 duplicated names shared by 1,730 counties across multiple states. GDIS struggled with this duplication. To illustrate the improvement, (Fig.~\ref{hurricane_irma}) contrasts GDIS and Geo-Disasters geocoding for Hurricane Irma, which made landfall in Florida in September 2017, alongside the USGS water-extent as an external reference \cite{USGS_IrmaWaterFootprint}.

%In addition to correcting outdated or nonexistent ISO codes, the dataset’s geometries were simplified to facilitate easier handling and analysis. These refinements improve data consistency, accuracy, and usability for spatial analysis and disaster impact assessment.

%The Methods should include detailed text describing any steps or procedures used in producing the data, including full descriptions of the experimental design, data acquisition assays, and any computational processing (e.g. normalization, image feature extraction). See the detailed section in our submission guidelines for advice on writing a transparent and reproducible methods section. Related methods should be grouped under corresponding subheadings where possible, and methods should be described in enough detail to allow other researchers to interpret and repeat, if required, the full study. Specific data outputs should be explicitly referenced via data citation (see Data Records and Citing Data, below).

%Authors should cite previous descriptions of the methods under use, but ideally the method descriptions should be complete enough for others to understand and reproduce the methods and processing steps without referring to associated publications. There is no limit to the length of the Methods section. Subheadings should not be numbered.

%\subsection*{Subsection}

%Example text under a subsection. Bulleted lists may be used where appropriate, e.g.

%\begin{itemize}
%\item First item
%\item Second item
%\end{itemize}

%\subsubsection*{Third-level section}
%Topical subheadings are allowed.

\section*{Data Records}

%The Data Records section should be used to explain each data record associated with this work, including the repository where this information is stored, and to provide an overview of the data files and their formats. Each external data record should be cited numerically in the text of this section, for example \cite{Hao:gidmaps:2014}, and included in the main reference list as described below. A data citation should also be placed in the subsection of the Methods containing the data-collection or analytical procedure(s) used to derive the corresponding record. Providing a direct link to the dataset may also be helpful to readers (\hyperlink{https://doi.org/10.6084/m9.figshare.853801}{https://doi.org/10.6084/m9.figshare.853801}).

%Tables should be used to support the data records, and should clearly indicate the samples and subjects (study inputs), their provenance, and the experimental manipulations performed on each (please see 'Tables' below). They should also specify the data output resulting from each data-collection or analytical step, should these form part of the archived record.

Geo-Disasters is openly hosted on Zenodo (\url{https://doi.org/10.5281/zenodo.15487667}), and the version-controlled code is available on figshare (\url{https://doi.org/10.6084/m9.figshare.29125907.v1}).
The dataset is distributed as two GeoPackage (.gpkg) files: 

\begin{itemize}
    \item Individual Subnational Locations (`disaster\_subnational\_90\_23.gpkg`): polygon features listing every affected sub-national unit for each event. The data columns are listed and described in Table \ref{tab:variables}.
    \item Aggregated Event Extents (`disaster\_national\_90\_23.gpkg`): dissolved polygons representing the aggregated spatial footprint of each event.
\end{itemize}

Packaging the data in GeoPackage (.gpkg) ensures full compatibility with modern GIS software and facilitates seamless spatial analysis. To reduce file size while retaining cartographic fidelity, we simplified each polygon with GeoPandas' Douglas–Peucker implementation (GeoSeries.simplify) using a tolerance of 0.005 degrees ($\sim$ 550 m at the equator). All geoprocessing was performed in planar---not spherical---geometry using the WGS 84 coordinate reference system (EPSG 4326).

\textbf{Licence and reuse:} The event footprints and all boundary geometries in Geo-Disasters are spatial derivatives of the GAUL 2015 dataset (FAO, 2015). Consequently, they are released under the GAUL 2015 Data Licence (see \url{https://developers.google.com/earth-engine/datasets/catalog/DataLicenseGAUL2015.pdf}). The data may be used, copied, and redistributed for non-commercial purposes only; every reuse or redistribution must carry the attribution “Boundaries © FAO 2015, GAUL – Global Administrative Unit Layers”; commercial exploitation requires prior written permission from FAO.

The accompanying geocoding scripts are our own work and are released separately under the MIT licence (\url{https://doi.org/10.6084/m9.figshare.29125907.v1}). If users need a version of the dataset that is permissively licensed for commercial reuse, they should remove the GAUL-derived geometries and keep only the non-spatial event attributes.

\section*{Technical Validation}

%This section presents any experiments or analyses that are needed to support the technical quality of the dataset. This section may be supported by figures and tables, as needed. This is a required section; authors must present information justifying the reliability of their data.

\begin{figure}[h]%
\centering
\centerline{\includegraphics[scale=0.5]{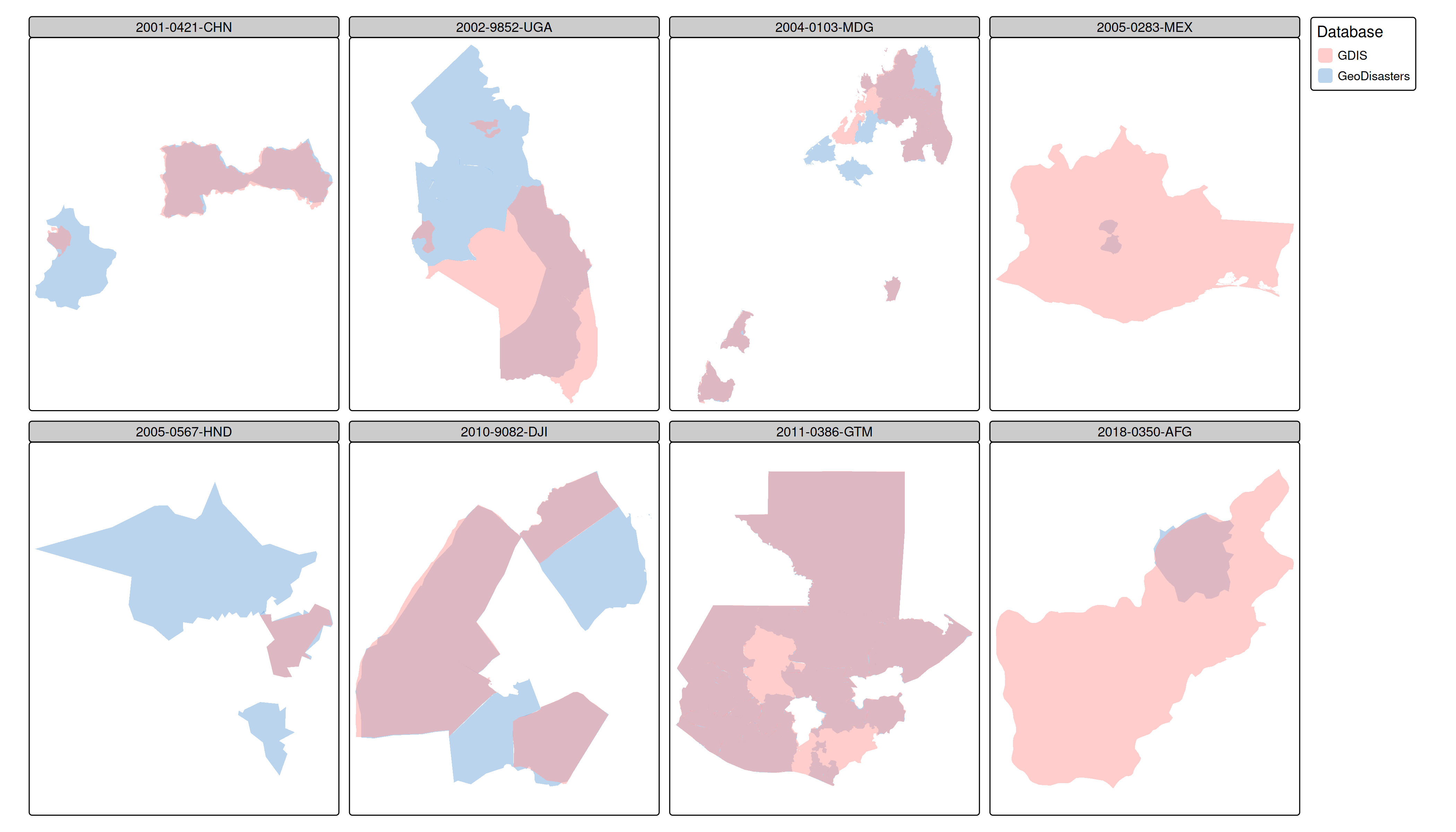}}
\caption{\textbf{A sample of disaster events overlap and mismatch from Geo-Disasters and the GDIS databases}. Extent, overlap and mismatch between randomly sampled disaster events from  Geo-Disasters and the GDIS databases. In the case of Geo-Disasters, we sampled only from events having the quality flag 1--highest quality where the GAUL information was provided by EM-DAT.}
\label{spatial_mismatch}
\end{figure}

Our validation approach consists of two main steps. First, we compare Geo-Disasters with GDIS to assess consistency. To ensure comparability, we examine event records from the overlapping period between both datasets, spanning 1990 - 2018, and consider only climate related disasters (flood, storm, extreme temperature, landslide, drought, and mass movement). During this period, GDIS records 7,218 events, while Geo-Disasters contains 7,546 events. However, only 5,620 events are common to both datasets. This gap arises from 52 incorrect ISO-3 country codes in GDIS, which generate incorrect disaster event identifiers. Next, to evaluate spatial accuracy, we assess the overlap between event geometries in both databases. If the spatial overlap of an event exceeds 90\% between both databases, the mismatch is considered negligible; otherwise, we classify it as a mismatch. The initial screening reveals that 2,785 events (49.5\%) show no mismatch, while 2,835 events (50.5\%) exhibit discrepancies.
Since we can identify events with the highest geocoding quality (Q = 1), we restrict the analysis to these cases. Among the 2,835 mismatched events, 2,166 carry the highest quality flag, with an average mismatch of 56.2\%. This high proportion underscores the inherent limitations of name-based geocoding. Such an approach introduces bias by creating discrepancies with the structured geocoding information provided by EM-DAT. To illustrate this issue, we present comparison plots of eight randomly selected events in both databases showing an example of spatial mismatch in different contexts (Fig.~\ref{spatial_mismatch}). These findings raise concerns about the reliability of GDIS, particularly because it lacks a systematic method for identifying uncertain geocoded events. Our quality-flag scheme in Geo-Disasters therefore allows users to filter out potentially unreliable locations and conduct more robust analyses.

\begin{figure}[h]%
\centering
\centerline{\includegraphics[scale=0.5]{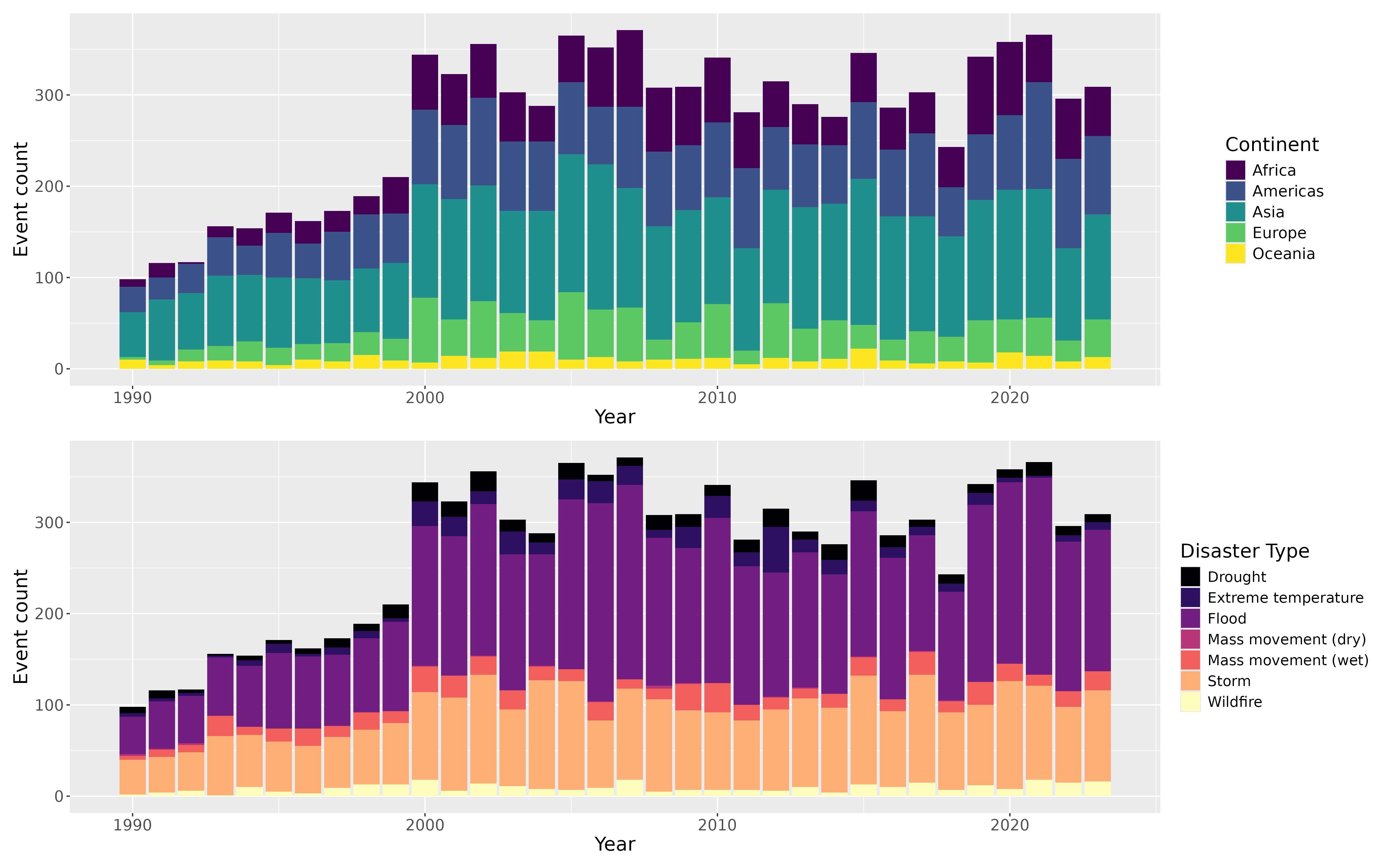}}
\caption{\textbf{Annual distribution of geocoded EM-DAT events by disaster type from 1990 to 2023}. The annual counts of the geocoded events follows a known pattern in EM-DAT, where the increase in the reporting quality stabilizes around the year 2000. Nevertheless, the event reporting in 1990's provides between 100 and 200 events per year of good quality, useful for studying disasters in the past three decades.}
\label{yearly_counts}
\end{figure}

\begin{figure}[h]%
\centering
\centerline{\includegraphics[scale=0.5]{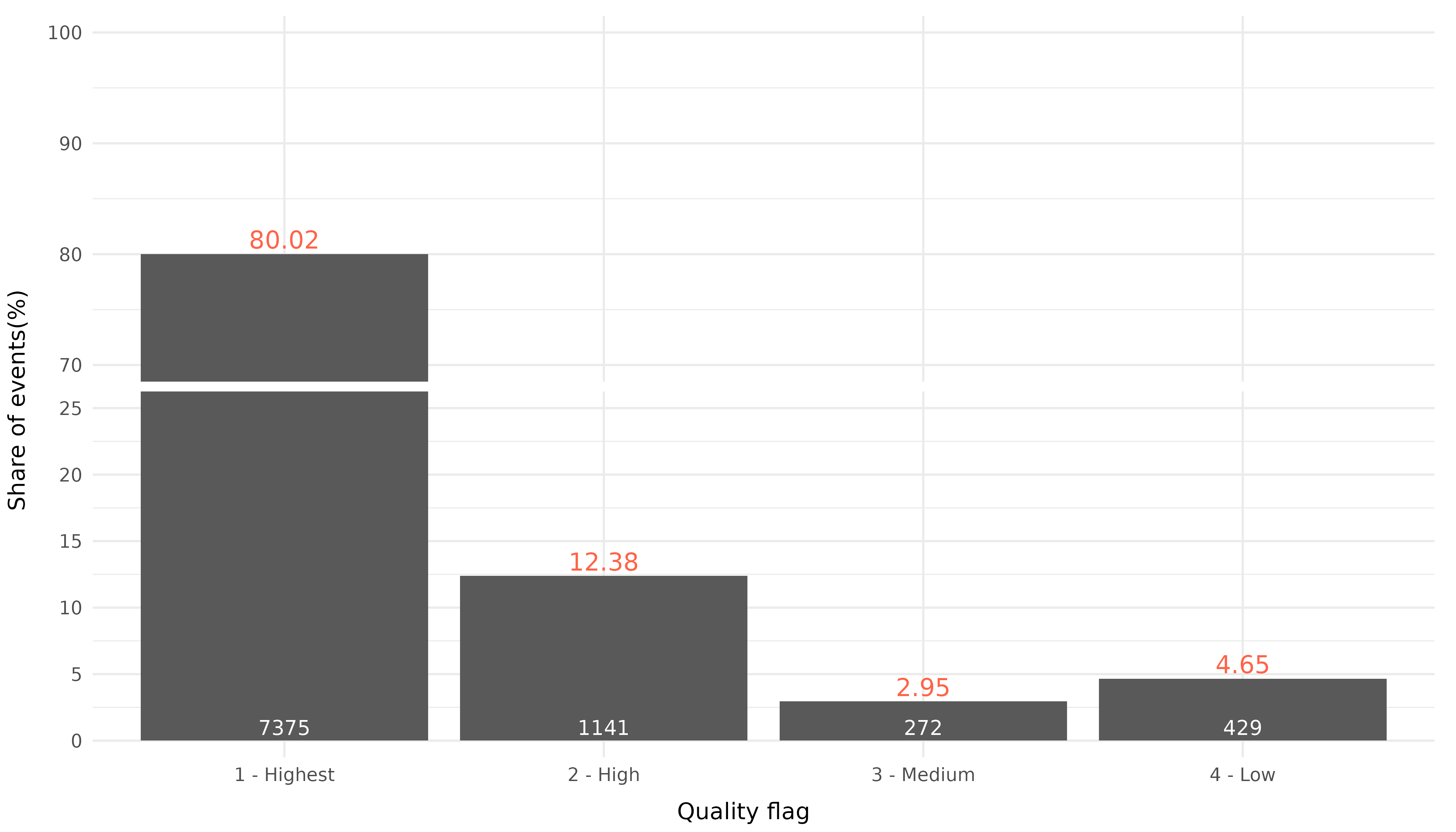}}
\caption{\textbf{Distribution of geocoding quality flags}. The quality flags are attributed at the location scale, and since an event is typically constituted of all impacted subnational locations in a given country, we find different combinations of the quality flags at the event scale. For 80\% of the events, the GAUL ID is provided for all locations that constitute the event. For the remaining 20\%, the locations are mostly identified by GeoNames then matched by names in 12.38 \%. Only in the remaining 7.62\% are the event locations identified with different approaches.}
\label{quality_flags}
\end{figure}

The second step of the validation focused on data quality over time and space and on the measures we adopted to improve the geocoding reliability. A well-known issue in EM-DAT is the gradual improvement in reporting quality over time, as the number of EM-DAT events reported per year stabilizes after 2000. Substantial geopolitical changes reshaped many states worldwide throughout the time span covered by EM-DAT. Most notably, the fall of the Iron Curtain and the collapse of the Soviet Union in 1991 redrew borders across eastern Europe and central Asia. Therefore, both factors influenced our decision to set 1990 as a cut-off point. In addition, meaningful analyses of historical disasters requires ancillary datasets--such as global land cover maps and human development indices--that are rarely available before 1990. Although the 1990-1999 period includes fewer events per year than the last two decades, the numbers remain comparable. These patterns are illustrated by Fig.~\ref{yearly_counts}, that shows a yearly breakdown of event counts by hazard type and continent.
Beyond temporal inconsistencies, disaster reporting also exhibits geographical biases that influence the dataset. The spatial distribution of geocoded events reveals distinct geographic patterns (Fig.~\ref{spatial_distribution}), suggesting potential biases in both disaster reporting and perception. For instance, heatwaves tend to be under-reported in lower-income countries, whereas droughts appear more frequently. This discrepancy may reflect variations in data availability, media attention, or differing perceptions of climate-related disasters. These findings highlight the importance of not only validating data quality over time but also understanding spatial reporting discrepancies, which can influence the interpretation of climate disaster trends.
To demonstrate the effectiveness of our approach, we examine the distribution of geocoding quality flags. Fig.~\ref{quality_flags} shows how events distribute across the four quality tiers after applying our worst-case rule. Approximately 80\% of all identified locations have the highest quality flag, while more than 92\% hold either the highest (1) or high-quality (2) flag. However, this does not necessarily imply that lower-quality locations are incorrect; rather, it indicates that we lack the means to cross-check and validate their consistency with EM-DAT reporting. GeoNames may have identified the correct location, but uncertainty remains. By publishing the flags, we enable users to choose the level of positional certainty appropriate to their application and indicate where community efforts could further enhance EM-DAT accuracy.

\begin{table}[h]
    \centering
    \renewcommand{\arraystretch}{1.2} % Adjust row height for better readability
    \begin{tabular}{|l|p{12cm}|}
        \hline
        \textbf{Variable Name} & \textbf{Description} \\
        \hline
        DisNo.        & Unique disaster identification number from EM-DAT. \\
        \hline
        ADM1\_NAME    & Name of the first-level GAUL administrative unit (province/state). \\
        \hline
        ADM1\_CODE    & GAUL code of the first-level unit. \\
        \hline
        ADM2\_NAME    & Name of the second-level GAUL administrative unit (district/county). \\
        \hline
        ADM2\_CODE    & GAUL code of the second-level unit. \\
        \hline
        Location      & Original location string reported by EM-DAT. \\
        \hline
        geoNames      & Matched GeoNames toponym. \\
        \hline
        Province      & ADM1 returned by GeoNames. \\
        \hline
        ISO           & ISO-3 country code. \\
        \hline
        admin\_level  & Assigned administrative level (1 = ADM1, 2 = ADM2). \\
        \hline
        geocoding\_q  & Geocoding quality flag (1 = highest, 4 = lowest). \\
        \hline
        geom          & Geometry (polygon, or multipolygon) representing the location. \\
        \hline
    \end{tabular}
    \caption{Description of Variables in the subnational location dataset}
    \label{tab:variables}
\end{table}
\FloatBarrier

\section*{Usage Notes}

%The Usage Notes should contain brief instructions to assist other researchers with reuse of the data. This may include discussion of software packages that are suitable for analysing the assay data files, suggested downstream processing steps (e.g. normalization, etc.), or tips for integrating or comparing the data records with other datasets. Authors are encouraged to provide code, programs or data-processing workflows if they may help others understand or use the data. Please see our code availability policy for advice on supplying custom code alongside Data Descriptor manuscripts.

%For studies involving privacy or safety controls on public access to the data, this section should describe in detail these controls, including how authors can apply to access the data, what criteria will be used to determine who may access the data, and any limitations on data use. 

The geocoded disasters can be linked to EM-DAT using the unique disaster identification number ("DisNo."). In addition, we supply the full geocoding framework and accompanying code, enabling users to update Geo-Disasters as new EM-DAT events are released. We also provide the R script used to generate the figures in the code repository.

\section*{Code availability}

%For all studies using custom code in the generation or processing of datasets, a statement must be included under the heading "Code availability", indicating whether and how the code can be accessed, including any restrictions to access. This section should also include information on the versions of any software used, if relevant, and any specific variables or parameters used to generate, test, or process the current dataset. 

 Version-controlled code is available on figshare (\url{https://doi.org/10.6084/m9.figshare.29125907.v1}).

\bibliography{sn-bibliography}

\section*{Acknowledgements} %(not compulsory)

%Acknowledgements should be brief, and should not include thanks to anonymous referees and editors, or effusive comments. Grant or contribution numbers may be acknowledged.
We thank the European Space Agency for funding ARCEME, Grant Number: 4000137109/22/I-EF and the European Commission for funding XAIDA, Grant Number: 101003469.\\
We thank Milena Moenks for her assistance in creating Fig. 1.

\section*{Author contributions statement}

Conceptualization: KT, MDM; Methodology: KT; Code: KT; Formal analysis: KT; Writing - Original Draft: KT; Writing - Review \& Editing: All authors.

\section*{Competing interests}

%The corresponding author is responsible for providing a \href{https://www.nature.com/sdata/policies/editorial-and-publishing-policies#competing}{competing interests statement} on behalf of all authors of the paper. This statement must be included in the submitted article file.

The authors declare no competing interests.

\section*{Supplementary information}

\section*{Figures \& Tables}

\FloatBarrier

\FloatBarrier

\FloatBarrier

\FloatBarrier

\FloatBarrier

\FloatBarrier

\end{document}